\documentclass[preprint,floats,aps,nofootinbib,amssymb]{revtex4}
\usepackage{mathrsfs}
\usepackage{slashed}
\usepackage{graphicx}
\usepackage{color}
\usepackage{dcolumn}
\usepackage{bm}
\usepackage{graphicx}
\usepackage{amssymb}
\usepackage{stackrel}

\begin{document}

\title{ Indirect detection imprint of leptophilic dark matter  }

\author{Wei Chao}
\email{chaowei@bnu.edu.cn}
\author{Zhuyao Wang}
\affiliation{$^1$Center for Advanced Quantum Studies, Department of Physics, Beijing Normal University, Beijing, 100875, China}

\vspace{3cm}

\begin{abstract}

In this paper we revisit constraints on the leptophilic dark matter (DM)  arising  from the DM indirect detection experiments. Interactions between the charged leptons and the scalar-type, Dirac-type or  vector-type DM are written in terms of effective operators. After classifying  all interactions that may  give non-zero signals in indirect detections, we study constraints on the parameter space of these effective interactions from the latest results of AMS-02, Planck, Fermi-LAT  and  H.E.S.S., as well as the observed relic abundance.  Main results are summarized in the Table. I.  It shows that the $\tau$-flavored DM is almost excluded by the DM indirect detection results in some scenario.

\end{abstract}

\maketitle

\section{Introduction}

Astrophysical observations have confirmed the existence of dark matter (DM), which can not be addressed within the minimal standard model (SM) of particle physics.  Although about 26.8\% of our universe is made by DM~~\cite{Aghanim:2018eyx}, one knows nothing about its nature except the gravitational effects, which catalyzed model buildings about the DM.  Of various DM models, the weakly interacting massive particle (WIMP)~\cite{Goldberg:1983nd,Ellis:1983ew,Jungman:1995df,Servant:2002aq,Cheng:2002ej,Bertone:2004pz} is a promising DM candidate, as it can naturally address the observed relic abundance in the framework of freeze-out with its mass around the electroweak scale and its interactions as weak as the weak nuclear force. The WIMP is also an important target of current and future DM detection experiments, which probe DM in the following ways: 
detecting nuclear recoils  in  underground laboratories induced by DM scattering off the nuclei (direct detections); reproducing DM particle by-hand in the large hadron collider (LHC) and detecting its signal in mono-X channel (direct detections); measuring the flux of  cosmic ray induced by the DM annihilation or decaies  with the help of satellite or ground-based telescopes (indirect detections). 

Up-to-now, no DM-like event has been observed  in underground direct detection experiments~\cite{Akerib:2013tjd,Tan:2016zwf,Aprile:2017iyp},  which in turn put constraints on the parameter space of some promising DM models. Improving the sensitivity of these experiments may enlarge the search scope, but these experiments will soon encounter the  background coming from the coherent neutrino-nuclei scattering, which may significantly reduce the sensitivity of direct detection experiments~\cite{Billard:2013qya,Chao:2019pyh}. From this point of view, indirect detections and collider searches are good complementarities to direct detections.  Notice that colliders are actually mediator machines, as a result  they are only powerful tools in studying some well-motivated ultraviolet-complete models that accommodate WIMP candidate in their mass spectrum.  Actually there are effective field theory (EFT) approaches to DM~\cite{Fitzpatrick:2012ix,Hertzberg:2012qn,Fedderke:2014wda,Ovanesyan:2014fwa,Krall:2014dba,Liem:2016xpm,Zheng:2010js,Yu:2011by,Chang:2013oia,DeSimone:2016fbz}, in which interactions between DM and the SM particles are described by high dimensional effective operators and there are only two free parameters: the DM mass, $m_{\rm DM}$ and the cutoff scale of the operator, $\Lambda$.  DM EFT has been proved to be a powerful tool in DM direct and indirect detections as the energy scale of these experiments lies far below $\Lambda$, however it will turns to be invalid in collider searches~\cite{Busoni:2013lha,Bauer:2016pug} to certain cutoff energy scale. 

In this paper, we study signals of leptophilic DM~\cite{Fox:2008kb,Bi:2009uj,Cohen:2009fz,Chao:2010mp,Chao:2012sz,Chao:2017yjg,Cai:2014hka} in indirect detection experiments. Assuming DM is scalar boson, Dirac fermion or vector boson, we write down the effective DM-charged lepton interactions in EFT approach, as well as the corresponding thermal averaged annihilation cross sections.  Then we study available parameter space of these interactions by taking into account the most recent results of  Planck~\cite{Aghanim:2018eyx}, AMS-02~\cite{Aguilar:2013qda}, Fermi-LAT~\cite{Ackermann:2015zua} and H.E.S.S.~\cite{Abdallah:2016ygi} experiments, as well as constraint of the observed relic abundance. Our main numerical results are summarized in the table.~\ref{results}. It shows that the $\tau$-lepton flavored DM is almost excluded by the indirect detection results in some scenario. These results may provide a guideline for DM model buildings. 

\begin{table}[t]
\centering
\begin{tabular}{c| c|c|c|c|c}
\hline
\hline  Flavors  &  ${\cal O}_{{\cal S}1},{\cal O}_{{\cal S}2}$ & ${\cal O}_{{\cal F}2},{\cal O}_{{\cal S}4}$ & ${\cal O}_{{\cal F}5},{\cal O}_{{\cal S}7}$ & ${\cal O}_{{\cal V}1},{\cal O}_{{\cal V}2}$  & ${\cal O}_{{\cal V}7}$\\
\hline
$e$ & $> 234~{\rm GeV}$ & $> 227~{\rm GeV}$ & $> 220~{\rm GeV}$ & $> 242~{\rm GeV}$ & $> 205~{\rm GeV}$ \\ 
\hline
$\mu$ & $> 162~{\rm GeV}$ & $> 145~{\rm GeV}$& $> 140~{\rm GeV}$& $> 142~{\rm GeV}$& $> 134~{\rm GeV}$ \\
\hline
$\tau$ &  $ (149, ~376)~{\rm GeV}$  &  $ (124, ~408)~{\rm GeV}$ &  $ (123, ~406)~{\rm GeV}$&  $ (120, ~419)~{\rm GeV}$&  $ (109, ~445)~{\rm GeV}$\\
& $\cup$ $>4.35 ~{\rm TeV}$& $\cup$ $>3.67~ {\rm TeV}$ & $\cup$ $>3.70 ~{\rm TeV}$ & $\cup$ $>3.47 ~{\rm TeV}$& $\cup$ $>3.11 ~{\rm TeV}$ \\
\hline
\hline
\end{tabular}
\caption{ Available DM mass range in various effective operators.   }\label{results}
\end{table}

The remaining of the paper is organized as follows: In section II we describe the  EFT that we will use in the analysis. In section III and IV we describe constraints of relic abundance and various indirect detection experiments. We present our numerical results in section V. The last part is concluding remarks.  

\section{effective operators}

We focus on interactions of DM with charged leptons. These interactions can only be efficiently probed by indirect detection experiments,  while their signals in direct detection experiments are loop suppressed~\cite{Chao:2019lhb,Chao:2017emq,Chao:2016lqd}.  In the limit where the particles  mediating the interaction between DM and leptons are heavier than the energies of interests, the interactions of the DM with the SM leptons can be written in terms of non-renormalizable effective operators 
\begin{eqnarray}
{\cal L}_i \sim  \zeta_i {\cal O }^i_{\rm L} {\cal O}^i_{\rm DM}
\end{eqnarray}
where $\zeta_i$ is the effective couplings that have dimensions of inverse mass up to the appropriate power, ${\cal O}_{\rm DM}^i$ are DM bilinears with ``i" representing their Lorentz structure, ${\cal O}_{\rm L}^i$ are gauge invariant lepton bilinears and the contraction of  ${\cal O}_{\rm DM}^i$ with ${\cal O}_{\rm L}^i$ results in Lorentz invariant effective operators. 

Assuming DM are complex scalar $\phi$, Dirac fermion $\chi$, or complex vector boson $X_\mu$, we list in the table~\ref{anomaly} all effective interactions. For the complex scalar, the DM bilinear $\phi^\dagger\phi$ couples to scalar type and pseudo-scalar type lepton bilinears,  while $\phi^\dagger i \stackrel{\leftrightarrow}{\partial_\mu} \phi$ may couple to vector type and axial-vector type lepton bilinears, resulting in four effective interactions, ${\cal O}_{Si}$ $(i=1,2,3,4)$.  For the Dirac DM, there are ten effective interactions ${\cal O}_{Fj}$ $(j=1,\cdots,10)$. 
For vector boson DM, there are eight effective operators ${\cal O}_{Vk}$ $(k=1,\cdots,8)$. Phenomenological implications of  these effective operators have been studied in many references, but it still make sense for us to revisit constraints on parameter spaces of these effective interactions by  updated results of indirect detections.

\begin{table}[t]
\centering
\begin{tabular}{l | l | l}
\hline
\hline ~~~Scalar DM  &  ~~~~~Dirac DM & ~~~~~Vector DM \\
\hline
${\cal O}_{ {\cal S} 1}^{}:~~ \zeta_{S1}^{} \phi^\dagger \phi \bar f f$ & ${\cal O}_{ {\cal F} 1}^{}:~~ \zeta_{F1}^{} \bar \chi \chi  \bar ff$ & ${\cal O}_{ {\cal V} 1}^{}:~~ \zeta_{V1}^{} X_\mu^{*} X^\mu \bar ff$ \\
& ${\cal O}_{ {\cal F} 2}^{}:~~ \zeta_{F2}^{} \bar \chi i \gamma^5 \chi \bar f f$  & \\
\hline
${\cal O}_{ {\cal S} 2}^{}:~~ \zeta_{S2}^{} \phi^\dagger \phi \bar f i \gamma_5^{} f$ &   ${\cal O}_{ {\cal F} 3}^{}:~~ \zeta_{F3}^{} \bar \chi \chi  \bar f  i \gamma^5 f$  & ${\cal O}_{ {\cal V} 2}^{}:~~ \zeta_{V2}^{} X_\mu^{*} X^\mu \bar f i \gamma_5^{} f$ \\
& ${\cal O}_{ {\cal F} 4}^{}:~~ \zeta_{F3}^{} \bar \chi i \gamma^5 \chi  \bar f  i \gamma^5 f$  &\\
\hline
${\cal O}_{ {\cal S} 3}^{}:~~ \zeta_{S3}^{} (\phi^\dagger i \stackrel{\leftrightarrow}{\partial_\mu} \phi ) \bar f \gamma^\mu f $  &  ${\cal O}_{ {\cal F} 5 }^{}:~~ \zeta_{F5}^{} \bar \chi \gamma^\mu  \chi  \bar f  \gamma_\mu f$& ${\cal O}_{ {\cal V} 3}^{}:~~ \zeta_{V3}^{} X_\rho^{*} i \stackrel{\leftrightarrow}{\partial_\mu} X^\rho \bar f \gamma_\mu^{}  f$  \\
& ${\cal O}_{ {\cal F} 6 }^{}:~~ \zeta_{F6}^{} \bar \chi \gamma^\mu  \gamma^5  \chi  \bar f  \gamma_\mu f$ & ${\cal O}_{ {\cal V} 4}^{}:~~ \zeta_{V4}^{} \varepsilon^{\mu \nu \rho \sigma} X_\mu^{*} i \stackrel{\leftrightarrow}{\partial_\nu} X_\rho \bar f \gamma_{\sigma}  f$ \\
\hline
${\cal O}_{ {\cal S} 4}^{}:~~ \zeta_{S4}^{} (\phi^\dagger i \stackrel{\leftrightarrow}{\partial_\mu} \phi ) \bar f \gamma^\mu \gamma^5 f $  &  ${\cal O}_{ {\cal F} 7 }^{}:~~ \zeta_{F7}^{} \bar \chi \gamma^\mu   \chi  \bar f  \gamma_\mu \gamma^5 f$ & ${\cal O}_{ {\cal V} 5}^{}:~~ \zeta_{V5}^{} X_\rho^{*} i \stackrel{\leftrightarrow}{\partial_\mu} X^\rho \bar f \gamma_\mu^{}  \gamma_5^{} f$ \\
& ${\cal O}_{ {\cal F} 8 }^{}:~~ \zeta_{F8}^{} \bar \chi \gamma^\mu  \gamma^5  \chi  \bar f  \gamma_\mu \gamma^5 f$ & ${\cal O}_{ {\cal V} 6 }^{}:~~ \zeta_{V6}^{} \varepsilon^{\mu \nu \rho \sigma} X_\mu^{*} i \stackrel{\leftrightarrow}{\partial_\nu} X_\rho \bar f \gamma_{\sigma} \gamma_5  f$ \\
\hline 
&${\cal O}_{ {\cal F} 9 }^{}:~~ \zeta_{F9}^{} \bar \chi \sigma^{\mu \nu } \chi  \bar f  \sigma_{\mu \nu } f$& ${\cal O}_{ {\cal V} 7 }^{}:~~ \zeta_{V7}^{} (X_\mu^* X_\nu^{} -X_\nu^* X_\mu^{} ) \bar f \sigma_{\mu \nu}  f$\\
&${\cal O}_{ {\cal F} 0 }^{}:~~ \zeta_{F0}^{} \varepsilon^{\mu\nu\rho \sigma} \bar \chi \sigma_{\mu \nu } \chi  \bar f  \sigma_{\rho \sigma } f$& ${\cal O}_{ {\cal V} 8 }^{}:~~ \zeta_{V8}^{} \varepsilon^{\mu\nu \rho \sigma} (X_\mu^* X_\nu^{} -X_\nu^* X_\mu^{} ) \bar f \sigma_{\rho \sigma}  f$\\
\hline
\hline
\end{tabular}
\caption{ The effective DM-lepton interactions, where $f=e,~\mu,~\tau$.  }\label{anomaly}
\end{table}

\begin{table}[h]
\centering
\begin{tabular}{c l|cl|cl}
\hline
\hline & Scalar DM  &  & Fermionic DM &  & Vector DM \\
\hline
~~~~~~~~& ~~~$\langle \sigma v\rangle$ & ~~~~~~~~& ~~~$\langle \sigma v\rangle$ & ~~~~~~~~& ~~~$\langle \sigma v\rangle$ \\
\hline
${\cal O}_{ {\cal S} 1}^{}:$ & ${1 \over 4 \pi }\zeta_{S1}^{2}  \left( 1- {3 \over 2 } x^{-1} \right) $ & ${\cal O}_{ {\cal F} 1}^{}:$ & ${ 1 \over 8 \pi  } \zeta_{F1}^2  m_\chi^2 x^{-1} $  &  ${\cal O}_{{\cal V}1}:$&  ${ 1 \over 12 \pi} \zeta_{V1}^2 \left( 1+{1\over 2 } x^{-1} \right)$\\
\hline
${\cal O}_{ {\cal S} 2}^{}:$ & ${1 \over 4 \pi } \zeta_{S2}^{2} \left( 1- {3 \over 2 } x^{-1} \right) $ & ${\cal O}_{ {\cal F} 2}^{}:$ & ${ 1  \over 2 \pi } \zeta_{F2}^2  m_\chi^2$ &  ${\cal O}_{{\cal V}2}:$& ${1 \over 12 \pi}  \zeta_{V2}^2  \left( 1+{1\over 2 } x^{-1} \right)$ \\
\hline
${\cal O}_{ {\cal S} 3}^{}:$ & ${1 \over  \pi  } \zeta_{S3}^{2 }  m_\phi^2 x^{-1}   $ & ${\cal O}_{ {\cal F} 3}^{}:$ & ${1 \over 8 \pi  }  \zeta_{F3}^2  m_\chi^2 x^{-1}$  &  ${\cal O}_{{\cal V}3:}$& ${1 \over 3 \pi }\zeta_{V3}^2  m_V^2 x^{-1}$ \\
\hline
${\cal O}_{ {\cal S} 4}^{}:$ & ${1  \over  \pi  } \zeta_{S4}^{2 } m_\phi^2 x^{-1} $ & ${\cal O}_{ {\cal F} 4}^{}:$ & ${ 1 \over 2 \pi } \zeta_{F4}^2 m_\chi^2$  &  ${\cal O}_{{\cal V}4}:$& ${10 \over 9 \pi } \zeta_{V4}^2 m_V^2 x^{-2}$ \\
\hline
&&${\cal O}_{ {\cal F} 5}^{}:$ & ${ 1  \over  \pi } \zeta_{F5}^2  m_\chi^2\left( 1- {1 \over 12} x^{-1} \right)$& ${\cal O}_{{\cal V}5}:$ & ${1\over 3 \pi} \zeta_{V5}^2 m_V^2 x^{-1}$\\
\hline
&&${\cal O}_{ {\cal F} 6}^{}:$ & ${ 1 \over  6\pi } \zeta_{F6}^2 m_\chi^2 x^{-1}$&${\cal O}_{{\cal V}6}:$& 0\\
\hline
&&${\cal O}_{ {\cal F} 7}^{}:$ & ${ 1 \over  \pi } \zeta_{F7}^2 m_\chi^2\left( 1-{1\over 12} x^{-1} \right)$&${\cal O}_{{\cal V}7}:$& ${1\over 9 \pi } \zeta_{V7}^2 \left( 2 + 3 x^{-1}\right) $\\
\hline
&&${\cal O}_{ {\cal F} 8}^{}:$ & $0$& ${\cal O}_{{\cal V}8 }:$ & ${1\over 9\pi } \zeta_{V8}^2 (8+12 x^{-1})$\\
\hline
&&${\cal O}_{ {\cal F} 9}^{}:$ & ${ 2 \over  \pi }  \zeta_{F9}^2 m_\chi^2 \left( 1- {1\over 12} x^{-1}\right)$&&\\
&&${\cal O}_{ {\cal F} 0}^{}:$ & ${ 8  \over  \pi }  \zeta_{F0}^2 m_\chi^2\left(1-{1\over 12 } x^{-1} \right) $&&\\
\hline
\hline
\end{tabular}
\caption{ Thermal averages of the reduced annihilation cross section for various effective DM-lepton interactions. We have neglected the lepton mass in the expression and $x^{-1} =\langle v^2 \rangle$ with $v$ the relative velocity in the laboratory frame.  }\label{reducedB}
\end{table}

\section{Relic density}
The cold DM in the early universe was in the local thermodynamic equilibrium. When its interaction rate drops below the expansion rate of the universe, the DM is said to be decoupled.  The evolution of the DM number density $n$, is governed by the Boltzmann equation~\cite{Gondolo:1990dk}:
\begin{eqnarray}
\dot{n} + 3 Hn =- \langle \sigma v_{\rm M\slashed{o}ller} \rangle ( n^2 -n_{\rm EQ}^2 ) \; , \label{boltzmann}
\end{eqnarray}  
where $H $ is the Hubble constant, $\sigma v_{\rm M\slashed{o}ller}$ is the total annihilation cross section multiplied by  the M$\slashed{\rm o}$ller velocity, $v_{\rm M\slashed{o}ller}=(|v_1 -v_2 |^2 -|v_1 \times v_2 |^2 )^{1/2}$, brackets denote thermal average and $n_{\rm EQ}$ is the number density at the thermal equilibrium. It has been shown that $\langle \sigma v_{\rm M\slashed{o}ller} =\langle \sigma v_{\rm lab} \rangle = 1/2 [1 + K_1^2 (x) /K_2^2 (x)] \langle \sigma v_{\rm cm} \rangle$~\cite{Gondolo:1990dk}, where $x=m/T$, $K_i$ is the modified Bessel functions of order $i$.

The present relic density of the DM is simply given by $\rho_\chi = M  n_\chi = M s_0 Y_{\infty} $,  where $s_0$ is the present entropy density. The relic density can finally be expressed in terms of the critical density~\cite{Bertone:2004pz}
\begin{eqnarray}
\Omega h^2 \approx 2 \times {1.07 \times 10^9 {\rm GeV}^{-1}  x_F \over M_{pl} \sqrt{g_*} ( a + 3 b/x_F )} \; ,
\end{eqnarray}
where $M_{pl}$ is the Planck mass,  $a$ and $b$, expressed in ${\rm GeV}^{-2}$, are the $s$-wave and the $p$-wave parts of the reduced annihilation cross section  and $g_*$ is the effective degrees of freedom at the freeze-out temperature $T_F$, $x_F= M/T_F $, which can be estimated through the iterative solution of the equation
\begin{eqnarray}
x_F = \ln \left[ c(c+2) \sqrt{45 \over 8 } {g\over 2 \pi^3 } { M_{dm}M_{pl} (a + 6 b/x_F ) \over \sqrt{g_* x_F }}\right] \; ,
\end{eqnarray}
where $c$ is a constant of order one determined by matching the late-time and early-time solutions.
It is conventional to write the relic density in terms of the Hubble parameter, $h=H_0 /100 {\rm km ~s^{-1}~Mpc^{-1}}$. Observationally, the DM relic abundance is determined to be $\Omega h^2 =0.1186\pm 0.0031$~\cite{Aghanim:2018eyx}.  Explicit expressions of $a$ and $b$  for various effective DM- lepton interactions are listed in the table.~\ref{reducedB}. Our results are agree with these presented in Refs.~\cite{Zheng:2010js,Yu:2011by}.

\section{Indirect detections}

In contrast to the direct detection of DM at underground laboratories, which measure the elastic DM-nuclei scattering cross section by blinding almost all cosmic rays, an indirect detection aims to look for hints of DM in cosmic rays on the background of   astrophysical sources. For review of indirect detections, we refer the reader to Refs.~\cite{Slatyer:2017sev,Hooper:2018kfv} and references cited therein. 
If DM is populated through interactions given in the Table.~\ref{anomaly}, it is possible that its annihilation rate into lepton pair is still large today. 
As is the case of ${\cal O}_{S1}$ and ${ \cal O}_{S2}$ for scalar DM, of ${\cal O}_{F2}$,  ${\cal O}_{F4}$, ${\cal O}_{F5}$, ${\cal O}_{F7}$, ${\cal O}_{F9}$ and ${\cal O}_{F0}$ for Dirac DM, and of ${\cal O}_{V1}$, ${\cal O}_{V2}$, ${\cal O}_{V7}$ and ${\cal O}_{V8}$ for vector DM, in which there are $s$-wave contribution to the reduced annihilation cross section as presented in the Table.~\ref{reducedB}. 
Annihilations into  charged lepton pairs can led to a diffuse emission of gamma rays through inverse Compton scattering and bremsstrahlung, which might be measured by Planck~\cite{Aghanim:2018eyx}, Fermi-LAT~\cite{Ackermann:2015zua} and H.E.S.S.~\cite{Abdallah:2016ygi}.
In addition the flux of cosmic ray can be recoded by the Fermi-LAT~\cite{Ackermann:2015zua}, AMS-02~\cite{Aguilar:2013qda} and DAMPE~\cite{TheDAMPE:2017dtc,Ambrosi:2017wek}. 
These measurements can lead to significant constraint on the annihilation rate today.
In this section, we will discuss these constraints separately.  

\begin{figure}[t]
\includegraphics[width=0.31\textwidth]{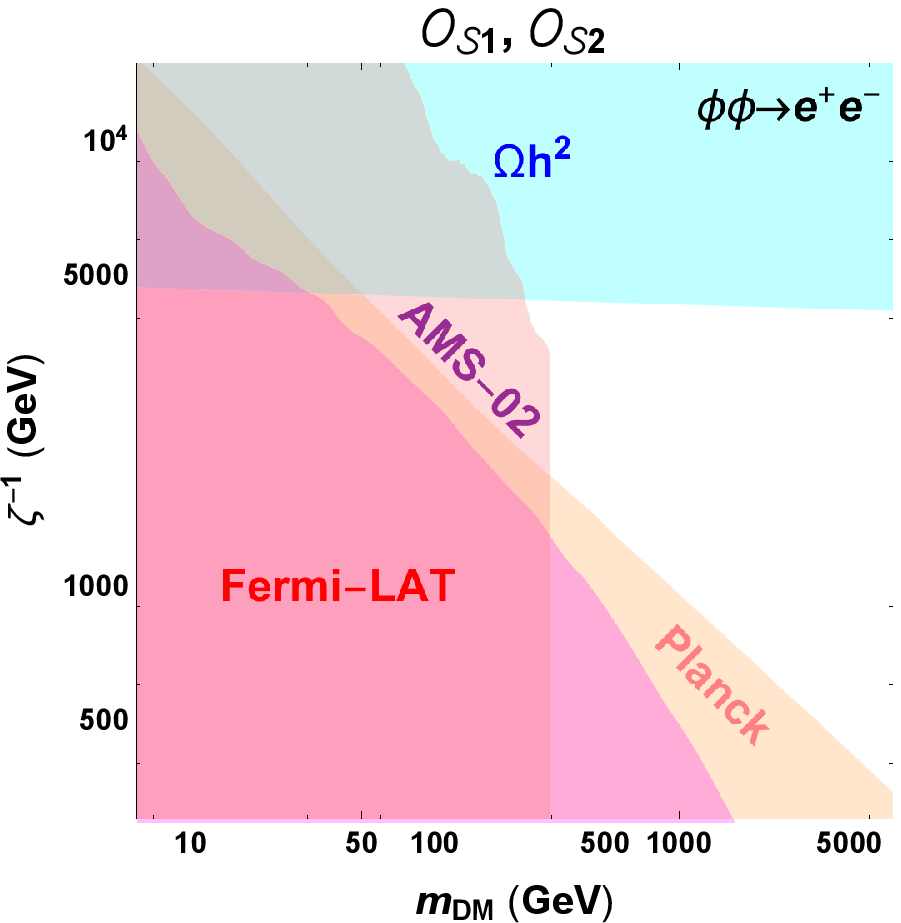}
\hspace{0.25cm}
\includegraphics[width=0.31\textwidth]{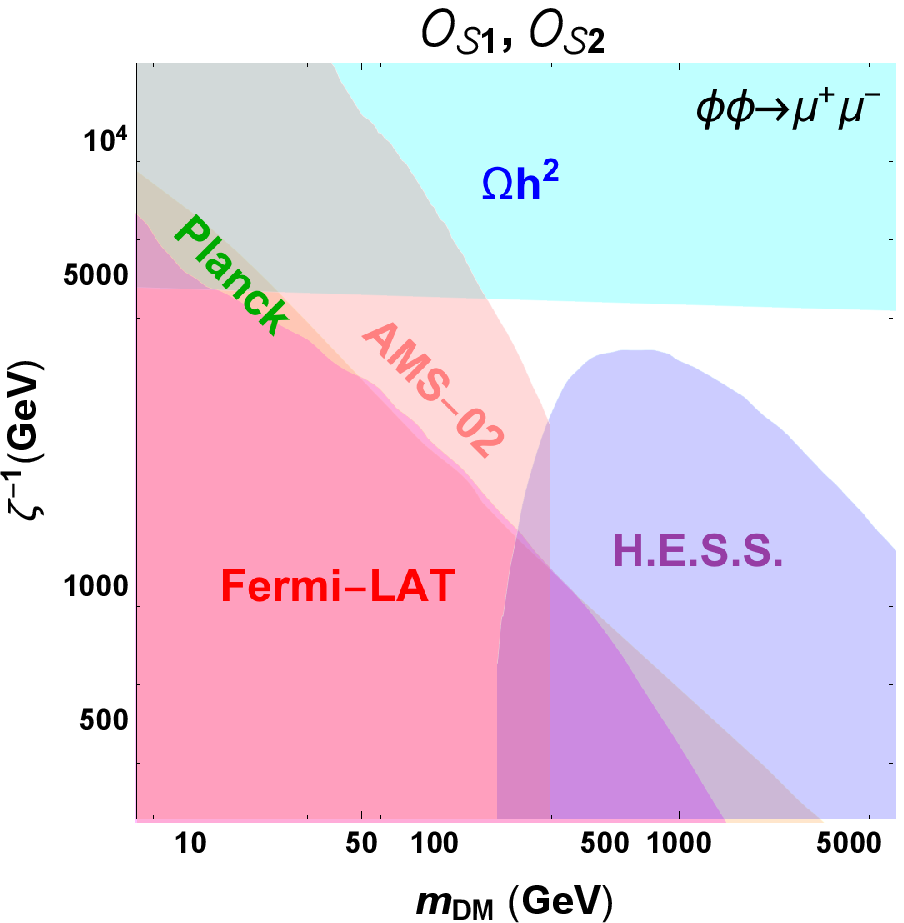}
\hspace{0.25cm}
\includegraphics[width=0.31\textwidth]{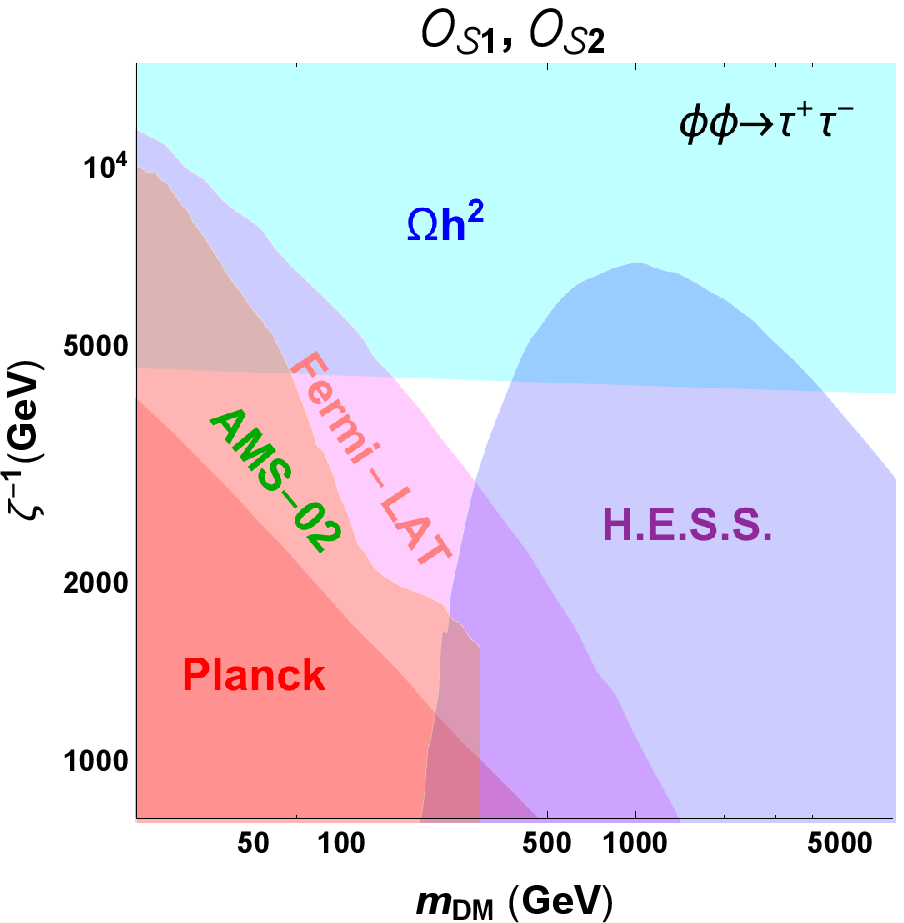}
\caption{ Constraints of operators ${\cal O}_{S1,S2}$ in the $m_{\rm DM}-\zeta^{-1}$ plane, with plots in the left, middle and right panels correspond to DM annihilating into $e^+e^-$, $\mu^+\mu^-$ and $\tau^+\tau^-$, respectively. Regions marked by light magenta, light orange, light red, purple and  light cyan are excluded by the Fermi-LAT, Planck, AMS-02, H.E.S.S. and the observed relic abundance, respectively.
} \label{fig:scalardm}
\end{figure}

\begin{figure}[t]
\includegraphics[width=0.31\textwidth]{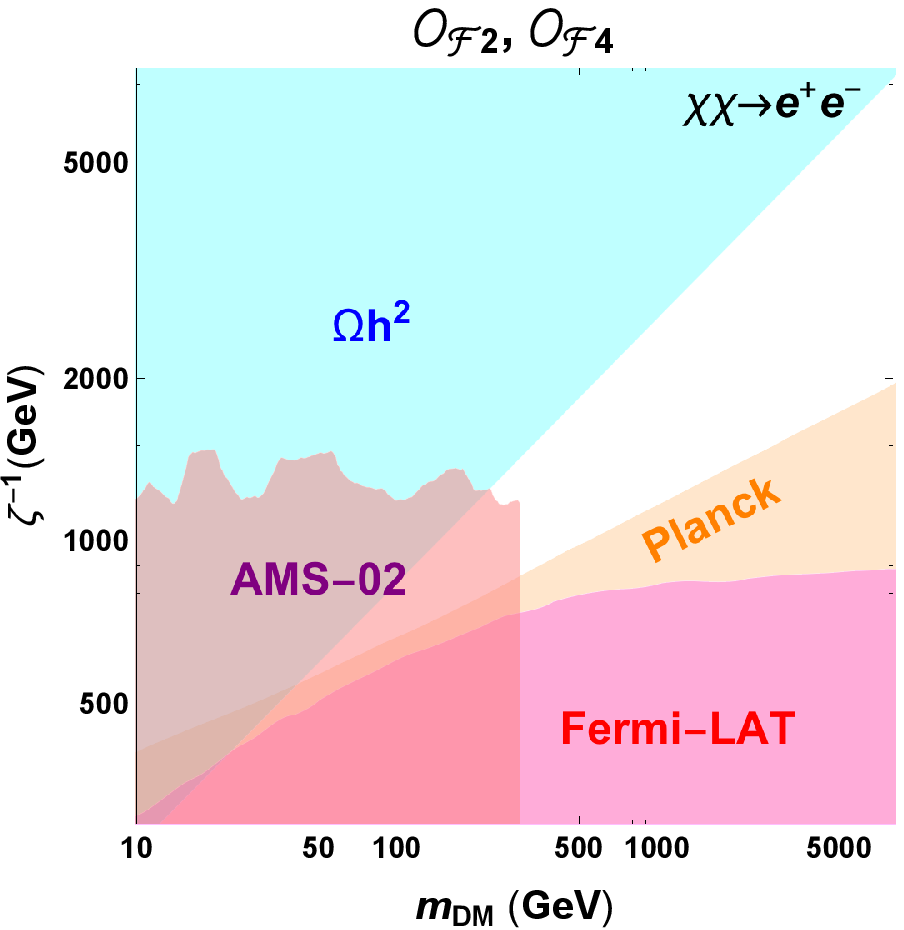}
\hspace{0.25cm}
\includegraphics[width=0.31\textwidth]{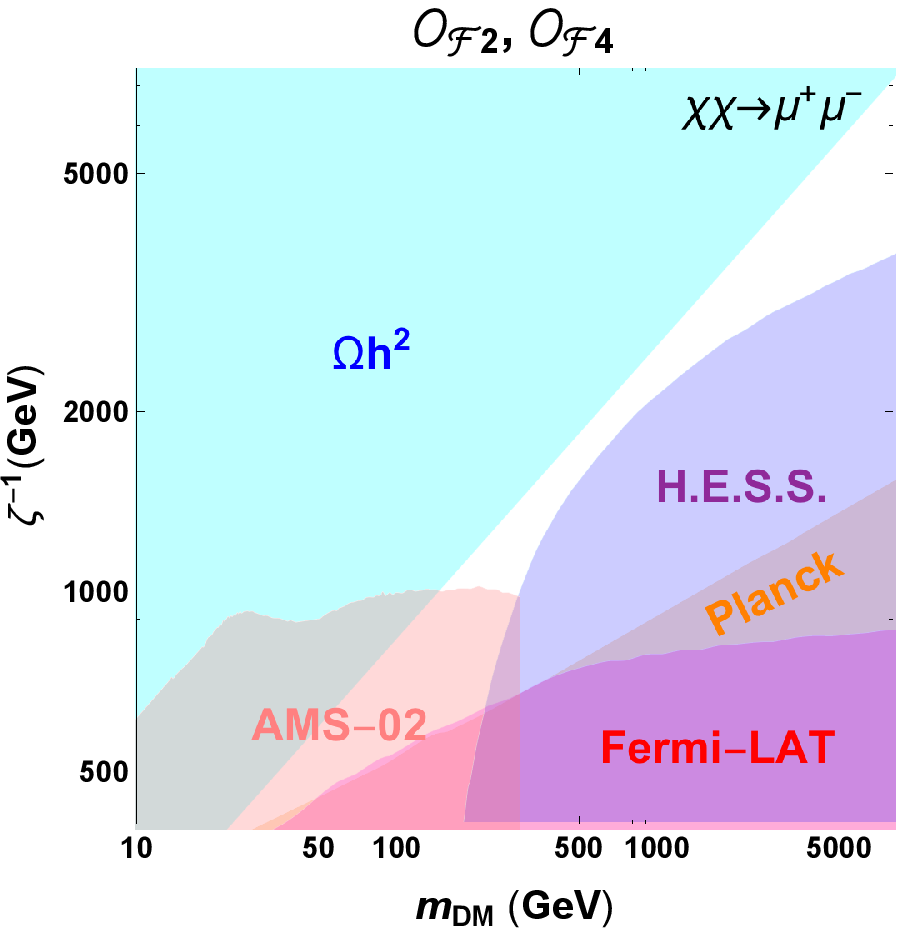}
\hspace{0.25cm}
\includegraphics[width=0.31\textwidth]{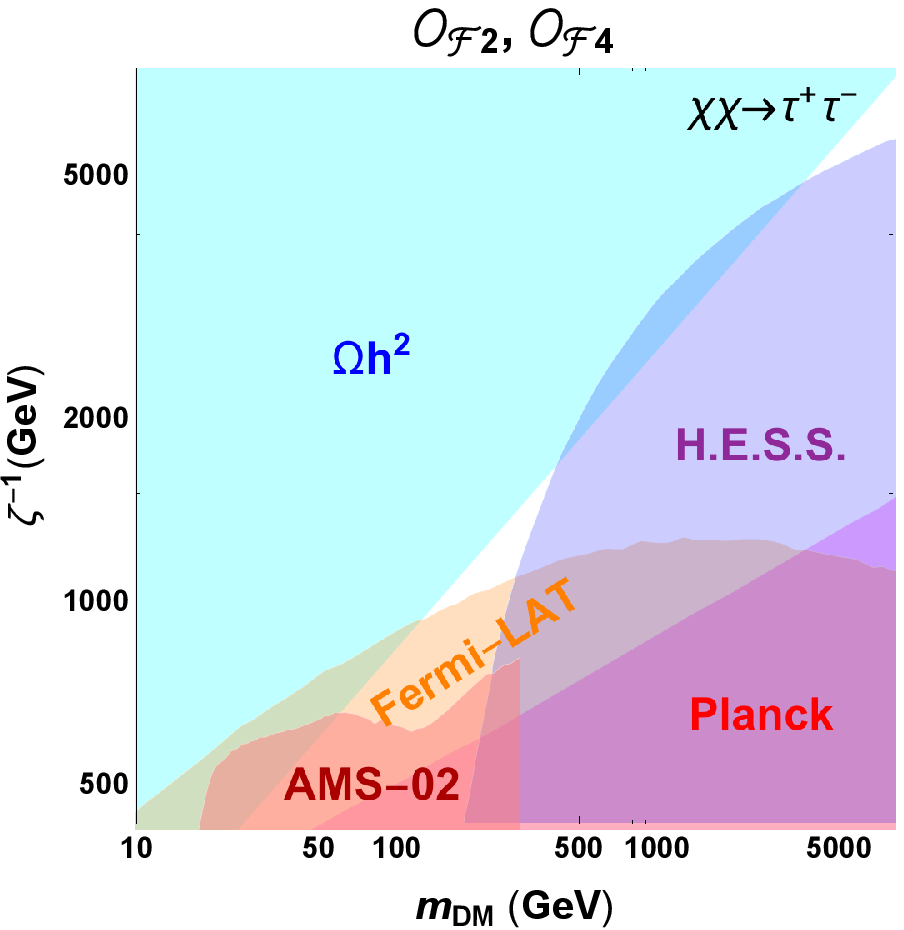}
\includegraphics[width=0.31\textwidth]{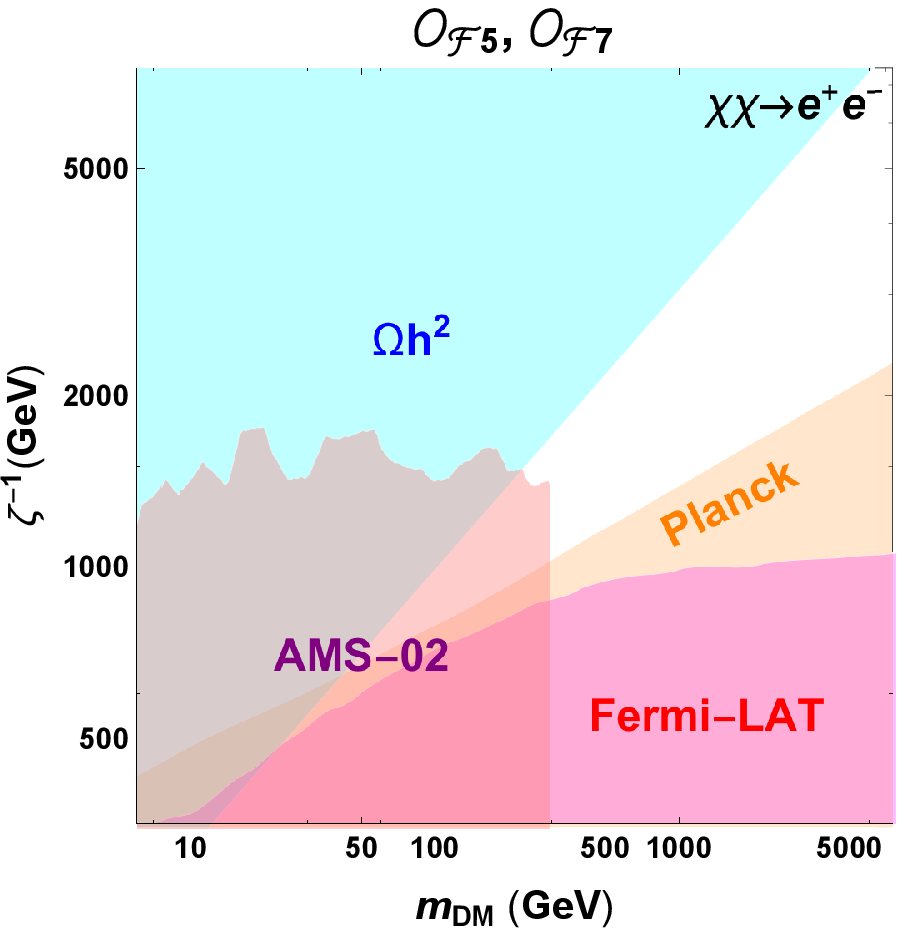}
\hspace{0.25cm}
\includegraphics[width=0.31\textwidth]{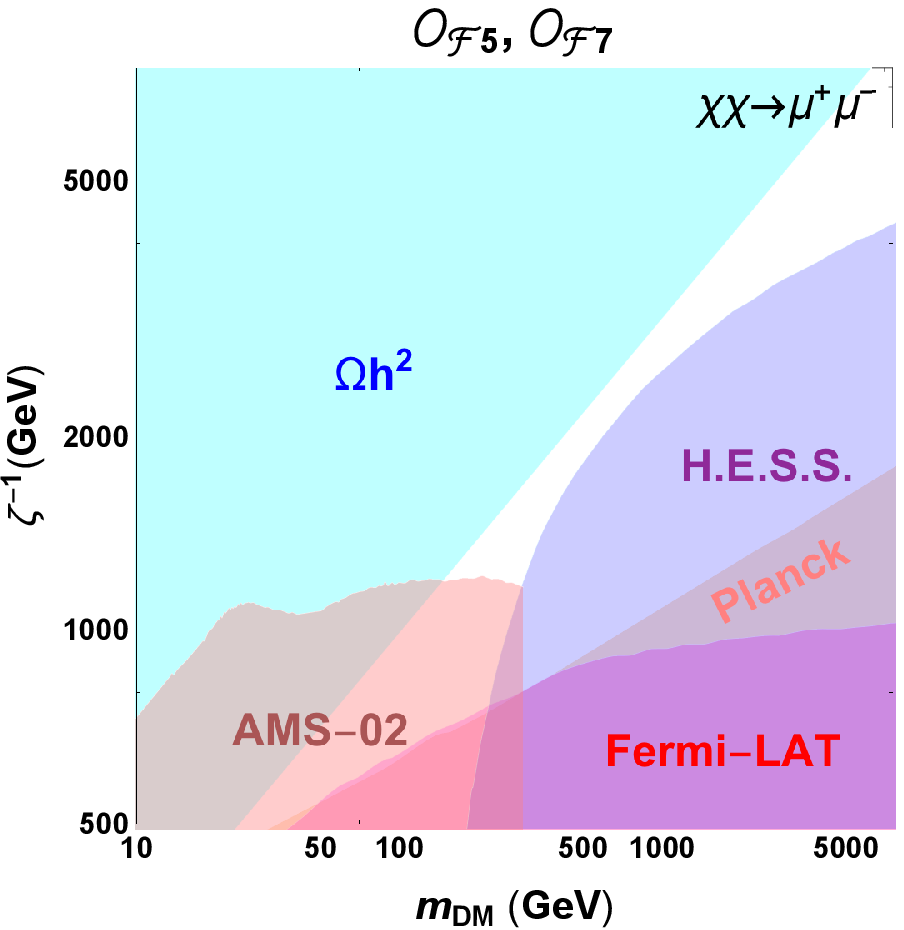}
\hspace{0.25cm}
\includegraphics[width=0.31\textwidth]{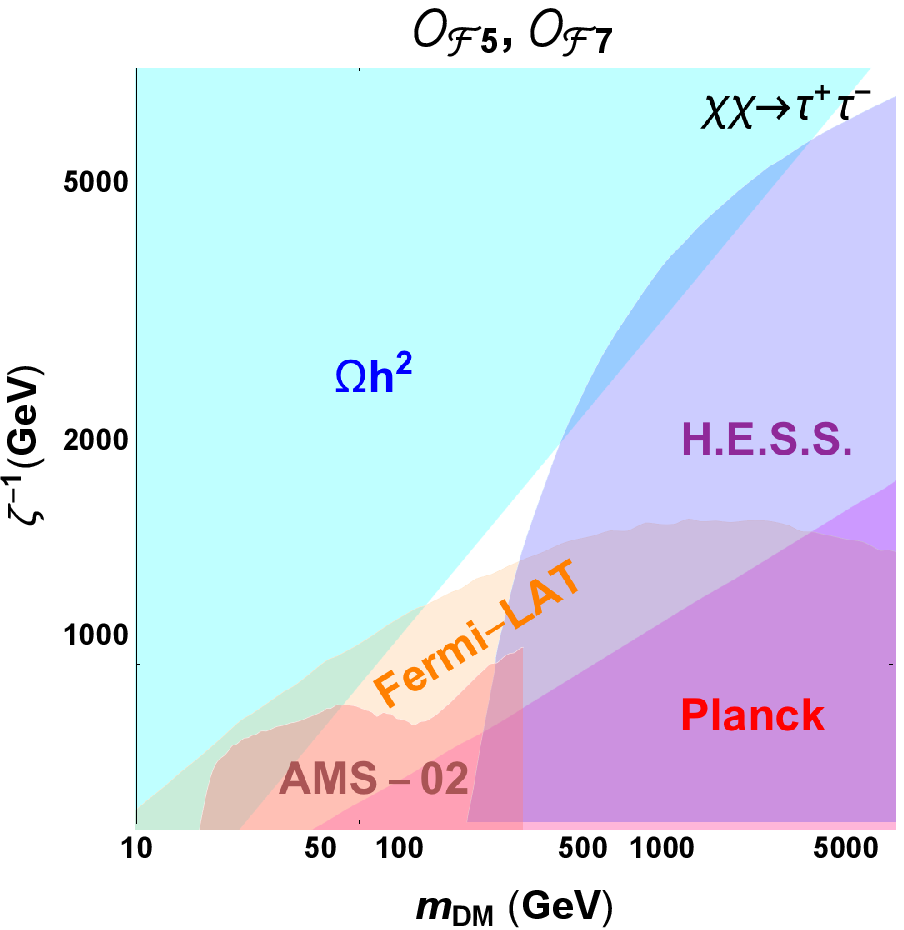}
\caption{ Constraints on the operators ${\cal O}_{{\cal F}2,{\cal F}4}$(first row) and ${\cal O}_{{\cal F}5,{\cal F}7}$(second row) in the $m_{\rm DM}-\zeta^{-1/2}$ plane, with plots in the left, middle and right correspond to DM annihilating into $e^+e^-$, $\mu^+\mu^-$ and $\tau^+\tau^-$ respectively. 
} \label{fig:diracdm}
\end{figure}

\begin{itemize}
\item Planck

Cosmic Microwave Background  (CMB) is powerful in probing DM annihilations into electrons.
DM annihilation in the dark ages, $z \leq 1200$, can cause additional ionization of the ambient hydrogen gas, perturbing the ionization history measured by the  Planck. 
The energy density deposit rate can be determined in term of the annihilation factor $p_{ann}^{}$ defined as
\begin{eqnarray}
p_{ann}^{} =f_{eff}^{} {\langle \sigma v\rangle \over m_{DM}} \; , \label{ann}
\end{eqnarray}
where $f_{eff}$ is a red-shift dependent efficiency function considering that injected energy is not equal to the one deposited in the intergalactic medium~\cite{Slatyer:2015jla}, $\langle \sigma v\rangle$ is the thermal average of the reduced DM annihilation cross section.
From Eq. (\ref{ann}), one can conclude that the precise measurements of CMB allows us to estimate the strength of DM interactions. 
Planck places an upper limit on $p_{ann}^{}$ at the $95\%$ C.L.,  read as $p_{ann}^{} <4.1\times 10^{-28}~{\rm  cm^3 s^{-1} GeV^{-1}}$~\cite{Ade:2015xua}. 
It can thus be used to constrain the $\langle \sigma v\rangle $. Limits are show in Figs.~\ref{fig:scalardm},~\ref{fig:diracdm} and \ref{fig:vectordm} marked in light-orange color. 

\item Fermi-LAT

$Fermi$ Large Area Telescope (LAT) is a gamma ray telescope which observes gamma rays emitted from the dwarf spheroidal satellite galaxies (dSphs) of the Milky Way that are supposed to be DM dominated. 
Fermi-LAT is sensitive to energies ranging from 20~{\rm MeV} to $>$  300 GeV. 
A combined analysis of 15 Milky Way dSphs using gamma ray data with energies between 500 MeV$\sim$ 500 GeV processed with the PASS-8 event level analysis shows that~\cite{Ackermann:2015zua}, no significant excess has been observed, which excluded the thermal relic annihilation $2.2\times 10^{-26} cm^3 s^{-1}$ for DM masses below $100~\text{GeV}$ annihilating through quark and $\tau$ lepton channels. 
The result of Fermi-LAT can constrain the effective DM-lepton interactions as shown in Figs.~\ref{fig:scalardm},~\ref{fig:diracdm} and \ref{fig:vectordm} marked by light magenta color. 

\item H.E.S.S.

SM particles could be produced from self-annihilation of WIMPs in high DM density regions  today. 
The High Energy Stereoscopic System (H.E.S.S.) array of ground based Cherenkov telescopes can detect very high energy  continuum gamma ray spectrum.
The differential gamma ray flux from the annihilation of WIMPs in a solid angle $\Delta \Omega$ can be written as
\begin{eqnarray}
{ d \Phi \over d E_\gamma^{} }  (E_\gamma^{}, \Delta \Omega) = { \langle \sigma v \rangle  \over 8 \pi m_{\rm WIMP}^2 } {d N_\gamma \over d E_\gamma } (E_\gamma^{} ) \times J(\Delta \Omega) 
\end{eqnarray}
where $\langle  \sigma v \rangle$ is the thermal average of the reduced annihilation cross section,  $dN_\gamma /dE_\gamma =\sum_f B_f  dN_\gamma^f /d E_\gamma^{} $ is the total  differential $\gamma$-ray from per annihilation ${\rm DM+DM}\to \bar f f (f^* f)$ with $B_f$ and $ dN_\gamma^f /d E_\gamma^{} $ the branching ratio of the annihilation and the differential $\gamma$-ray  emitted from $f$, $J(\Delta \Omega)$ is the integration of DM density square in a solid angle $\Delta \Omega$. 
The analysis to the data accumulated from 254 hours of Galactic center observations by H.E.S.S. shows that there are no significant $\gamma$-ray excess above the background. 
Upper limits on various reduced annihilation cross section were derived by assuming the Einasto~\cite{Springel:2008by} and Navarro-Frenk-White (NFW)~\cite{Navarro:1996gj} DM density profiles. 
The Constraints of the H.E.S.S. results on leptophilic  DM interactions are given in Figs.~\ref{fig:scalardm},~\ref{fig:diracdm} and \ref{fig:vectordm} marked by purple color.

\item AMS-02

DM annihilations in the Galactic Halo may give rise to the cosmic ray (CR) lepton spectrum, and its subsequent propagation can be described by the following diffusion equation~\cite{Vladimirov:2010aq},
\begin{eqnarray}
{\partial \psi\over \partial t} = \bigtriangledown\cdot(D\bigtriangledown \psi)+ {\partial (b\psi) \over \partial E} +Q
\end{eqnarray}
where $\psi$ is the differential number density of particles, $D$ is the diffusion coefficient, $b$ is the energy lose rate arising from inverse Compton scattering and synchrotron, $Q={1\over 4} \langle \sigma v\rangle (\rho_{\rm DM} /m_{\rm DM})^2 dN/dE$ being the injected lepton spectrum with $\rho_{\rm DM} $ the DM density and $m_{\rm DM}$ the DM mass. 

AMS-02 has measured the ratio of positrons to positrons-plus-electrons, which rises with respect to the ratio from secondary cosmic rays during CR propagation.  If interpreted as the annihilation of DM instead of astrophysical sources, the DM would be quite heavy. Even if the excess is not from the DM annihilation, it still can be used to constrain the annihilation of DM into leptons.   In our analysis, we use results of Refs.~\cite{Bergstrom:2013jra,Ibarra:2013zia} and  their relevant constraints are shown in Figs.~\ref{fig:scalardm},~\ref{fig:diracdm} and \ref{fig:vectordm} marked by light red color.

\end{itemize}

\section{Results}

In this section, we study constraints on the parameter space of various effective DM-lepton interactions using DM indirect detection results. Numerical results  for the scalar-type, Dirac-type and vector-type DMs are presented in the following:

\subsection{scalar DM}

As shown in the table. I, there are four types of interactions between the complex scalar DM and charged leptons, namely ${\cal O}_{S1,S2,S3,S4}$, of which only ${\cal O}_{S1}$ and ${\cal O}_{S2}$ have nonzero signals in indirect detection experiments since the s-wave component of the thermal averaged reduced annihilation cross section is zero for ${\cal O}_{S3}$ or ${\cal O}_{S4}$.  Moreover, the signals of ${\cal O}_{S1}$ and ${\cal O}_{S2}$ are the same. 

We show in the Fig.~\ref{fig:scalardm} constraints on the parameter space of operators ${\cal O}_{S1,S2}$ in the $m_{\rm DM}-\zeta^{-1}$ plane, where $m_{\rm DM}$ is the DM mass and $\zeta^{-1}~({\rm GeV})$ is the cut-off scale.  
Plots in the left, middle and right panels correspond to the DM annihilating into $e^+e^-$, $\mu^+\mu^-$ and $\tau^+\tau^-$ respectively.  In each plot, the regions marked by light magenta, light orange, light red, purple and  light cyan are excluded by the Fermi-LAT, Planck, AMS-02, H.E.S.S. and the observed relic abundance respectively, while regions marked by white color are still allowed.
For the annihilation channel into $e^+e^-$, the lower bound on the DM mass is about $234~{\rm GeV}$ and the corresponding upper bound on the cut-off scale is about $4422~{\rm GeV}$. In this case, constraints of relic abundance, AMS-02 and Planck are dominate, while that of H.E.S.S. is sub-dominate as the gamma ray yields over the $e^+e^-$ final state  is a secondary emission, which is expected from inverse Compton scattering of energetic electrons on ambient radiation fields. For the annihilation channel into $\mu^+\mu^-$, the constraint of H.E.S.S. turns out to be important for heavy DM, which gives a lower bound on the cut-off scale that depends on the DM mass as can be seen from the plot. In this case the lower bound on the DM mass is $162~{\rm GeV}$. For the annihilation channel into $\tau^+\tau^-$, constraints of relic abundance, Fermi-LAT and H.E.S.S. are dominate and the following mass region $(0, ~149~{\rm GeV})\cup(376~{\rm GeV},~4352~{\rm GeV})$ are excluded. 
The upper bound on the cut-off scale is about $4.4~{\rm TeV}$ for the DM mass range $(149~{\rm GeV},~376~{\rm GeV})$ and the lower bound on $\zeta^{-1}$ depends on the DM mass.

\subsection{Dirac DM}

There are ten effective interactions between the Dirac DM and charged leptons, of which ${\cal O_{F}}_2$, ${\cal O_{F}}_4$, ${\cal O_{F}}_5$, ${\cal O_{F}}_7$, ${\cal O_{F}}_9$ and ${\cal O_{F}}_{0}$ have non-zero signals in indirect detection experiments. 
We show in the Fig.~\ref{fig:diracdm} constraints on the parameter space of operators ${\cal O}_{{\cal F}2,{\cal F}4}$(first row) and ${\cal O}_{{\cal F}5,{\cal F}7}$(second row) in the $m_{\rm DM}-\zeta^{-1/2}$ plane, with plots in the left, middle and right panels correspond to DM annihilating into $e^+e^-$, $\mu^+\mu^-$ and $\tau^+\tau^-$ respectively. The lower bound on the DM mass in effective operators ${\cal O}_{{\cal F}2, {\cal F}4}$ is bout  $227~{\rm GeV}~(e^+e^-)$, $145~{\rm GeV}~(\mu^+\mu^-)$ and $124~{\rm GeV}~(\tau^+\tau^-)$. 
Furthermore, the DM mass region $(408~{\rm GeV},~3666~{\rm GeV})$ is excluded in the DM-$\tau$ lepton interactions. The lower bound on the DM mass in operators ${\cal O}_{{\cal F}5, {\cal F}7}$ is about $220~{\rm GeV}~(e^+e^-)$, $140~{\rm GeV}~(\mu^+\mu^-)$ and $123~{\rm GeV}~(\tau^+\tau^-)$, and the mass range $(406~{\rm GeV},~3698~{\rm GeV})$ is excluded in the DM-$\tau$ lepton interactions.  Notice that the thermal average of the annihilation cross sections in  ${\cal O_{F}}_9$ and ${\cal O_{F}}_{0}$ are similar to that in ${\cal O}_{{\cal F}5}$. As a result, constraints from indirect detection experiments are the same as these in ${\cal O}_{{\cal F}5}$ up to the following rescale $\sqrt{2} \zeta_{{\cal F}9}\to \zeta_{{\cal F}5}$ for ${\cal O_{F}}_9$ and  $2\sqrt{2} \zeta_{{\cal F}0}\to \zeta_{{\cal F}5}$ for ${\cal O_{{\cal F}}}_0$.

\begin{figure}[t]
\includegraphics[width=0.31\textwidth]{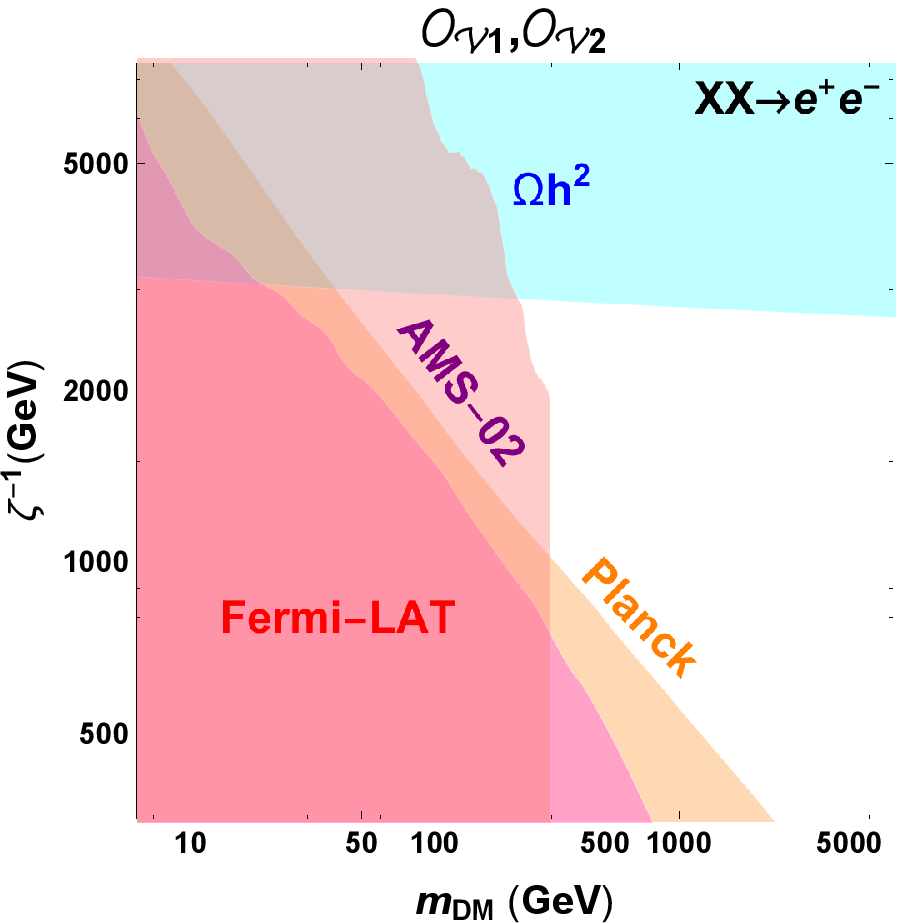}
\hspace{0.25cm}
\includegraphics[width=0.31\textwidth]{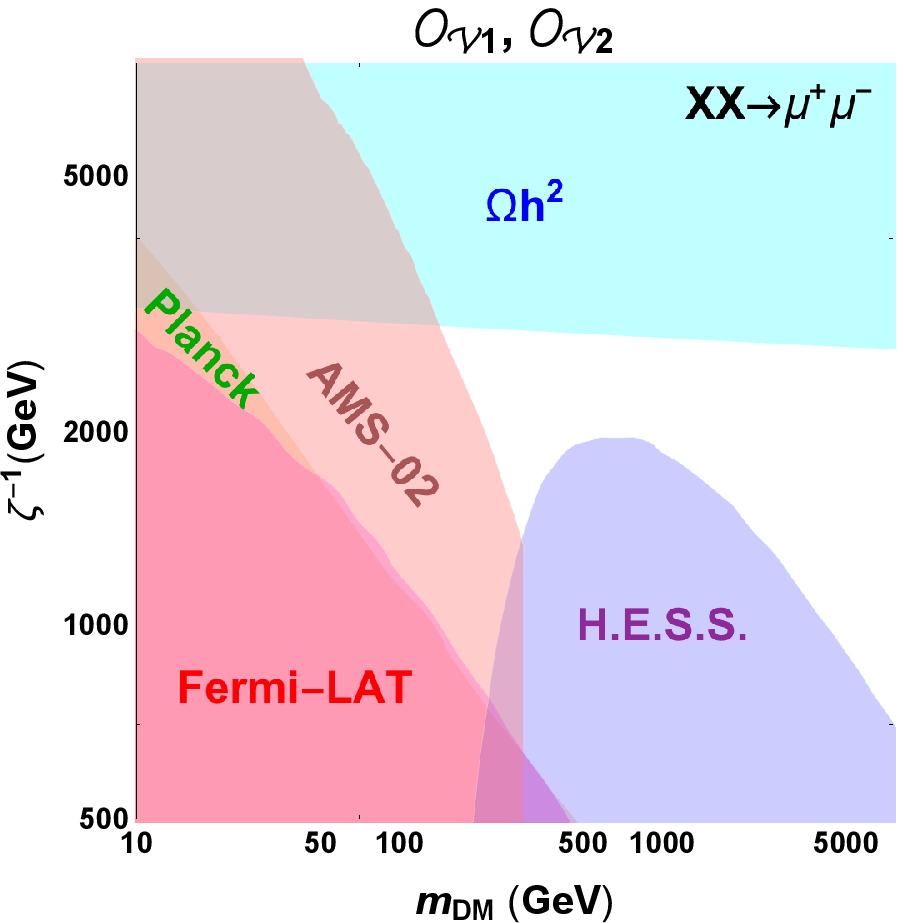}
\hspace{0.25cm}
\includegraphics[width=0.31\textwidth]{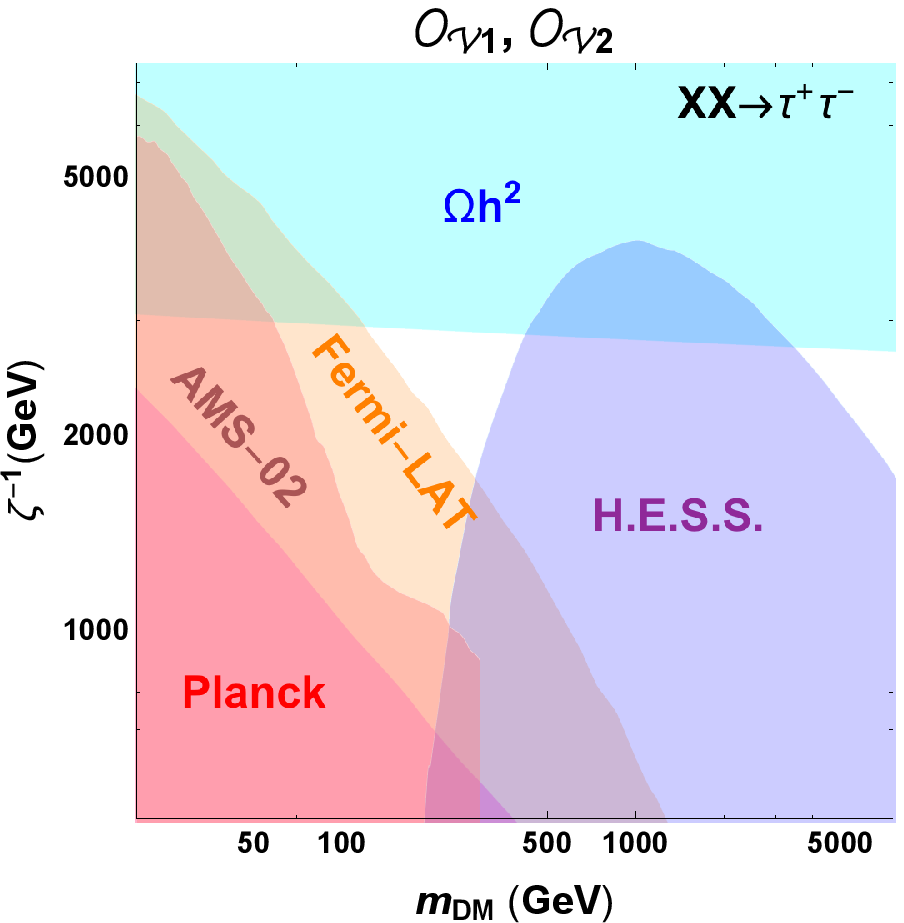}
\includegraphics[width=0.31\textwidth]{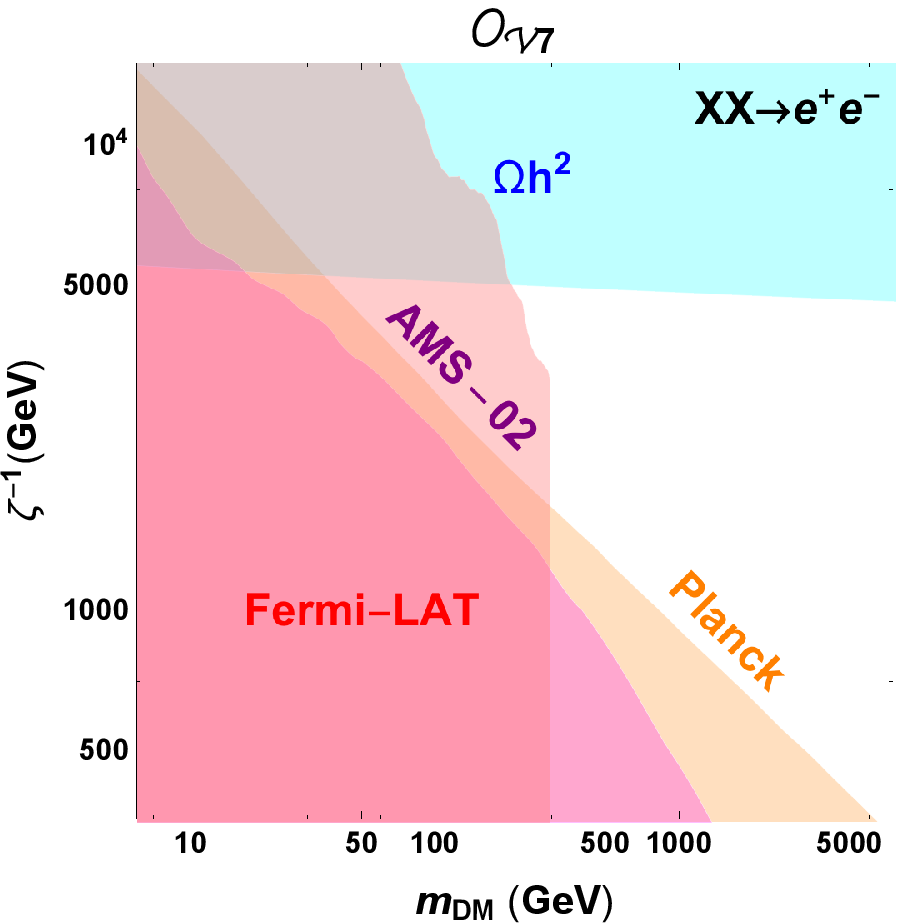}
\hspace{0.25cm}
\includegraphics[width=0.31\textwidth]{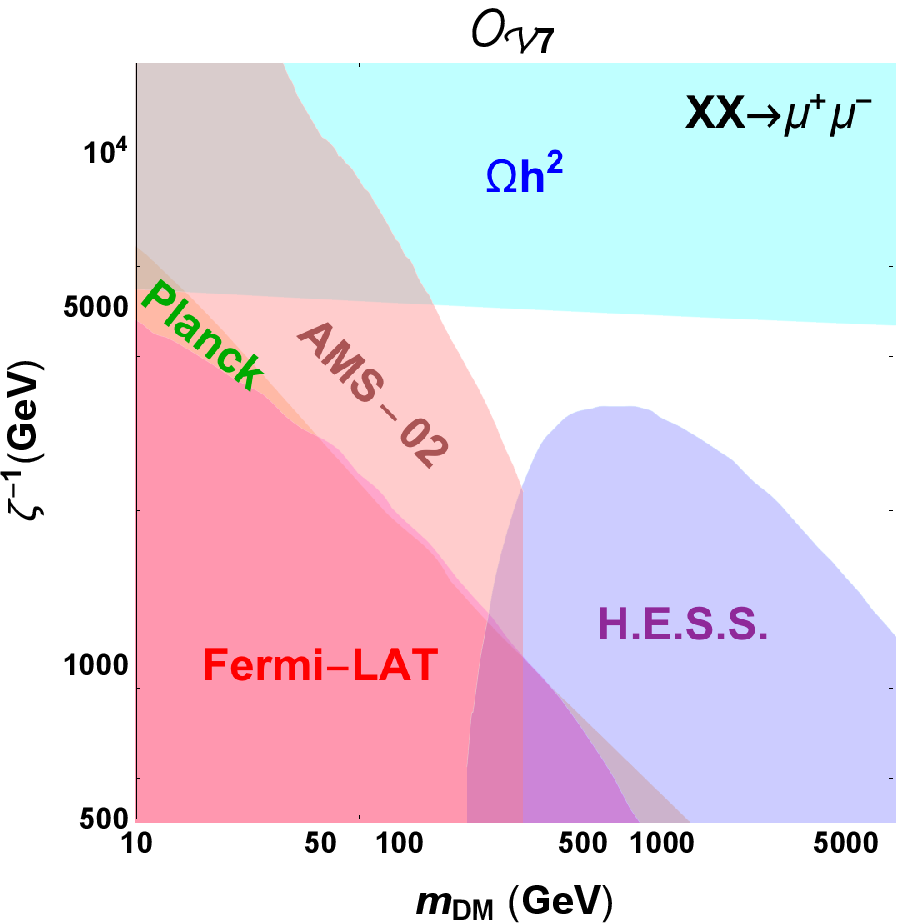}
\hspace{0.25cm}
\includegraphics[width=0.31\textwidth]{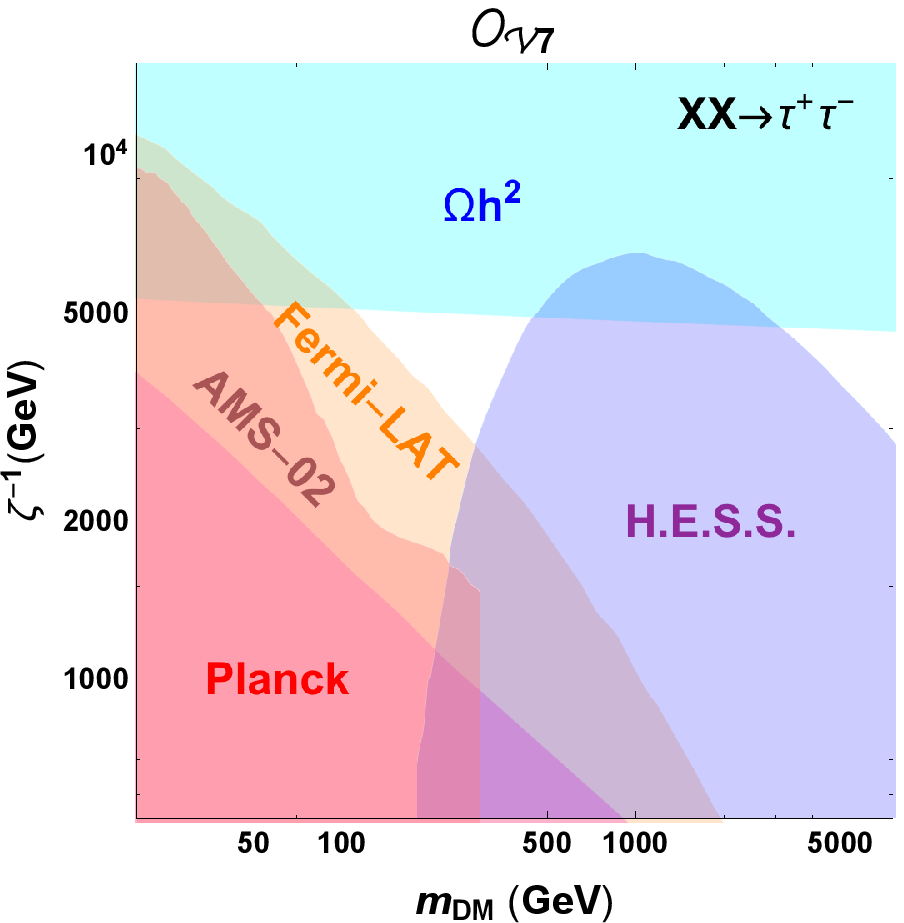}
\caption{ Constraints on the operators ${\cal O}_{{\cal V}1,{\cal V}2}$(first row) and ${\cal O}_{{\cal V}7}$(second row) in the $m_{\rm DM}-\zeta^{-1}$ plane, with plots in the left, middle and right panels correspond to DM annihilating into $e^+e^-$, $\mu^+\mu^-$ and $\tau^+\tau^-$ respectively. 
} \label{fig:vectordm}
\end{figure}

\subsection{Vector DM}

There are eight effective interactions between the charged leptons and the complex vector DM, of which ${\cal O}_{{\cal V}3,{\cal V}4,{\cal V}5,{\cal V}6}$ give null signal in indirect detection experiments, signals of ${\cal O}_{{\cal V}1}$ and ${\cal O}_{{\cal V}2}$ are the same, and the signal of ${\cal O}_{{\cal V}7}$ are similar to these of  ${\cal O}_{{\cal V}8}$ up to the rescale, $2\zeta_{{\cal V} 8 } \to \zeta_{{\cal V} 7}$. 

We show in the Fig.~\ref{fig:vectordm} constraints on the parameter space of operators ${\cal O}_{{\cal V}1,{\cal V}2}$(first row) and ${\cal O}_{{\cal V}7}$(second row) in the $m_{\rm DM}-\zeta^{-1}$ plane, with plots in the left, middle and right panels correspond to DM annihilating into $e^+e^-$, $\mu^+\mu^-$ and $\tau^+\tau^-$ respectively. The lower bounds on the DM mass in operators ${\cal O}_{{\cal V}1,{\cal V}2} ({\cal O}_{{\cal V}7})$ are $242(205)~{\rm GeV}$  in $e^+e^-$ channel, $142(134)~{\rm GeV}$  in $\mu^+\mu^-$ channel and $120(109)~{\rm GeV}$  in $\tau^+\tau^-$ channel. Furthermore, the following DM mass region $(419~{\rm GeV},~3471~{\rm GeV})$ and $(445~{\rm GeV},~3111~{\rm GeV})$ are excluded  by the combined constraints of the observed relic density and the H.E.S.S. results in the $\tau^+\tau^-$ channel for ${\cal O}_{{\cal V}1,{\cal V}2}$ and $ {\cal O}_{{\cal V}7}$ operators, respectively.

\section{Conclusion}

In this paper, we  systematically  studied    constraints on the leptophilic DM  arising from DM indirect detection experiments. Assuming the DM is complex scalar, Dirac fermion or complex vector boson, and it only interact with the charged leptons via the effective operators, we derived the lower bound of the DM mass  constrained by the relic abundance, AMS-02, Fermi-LAT, Planck and H.E.S.S.. Main numerical results are summarized in the Table.~\ref{results}.  
It shows that some effective interactions are strongly constrained by the indirect detection results.
This study may provide a guidance to the DM model buildings.

\begin{acknowledgments}
This work was supported by the National Natural Science Foundation of China  under Grant No. 11775025,  and  the Fundamental Research Funds for the Central Universities.
\end{acknowledgments}

\end{document}